\documentclass[dvips]{arxstspdf}
\usepackage{flushend}

\volume{24}
\issue{3}
\pubyear{2009}
\firstpage{294}
\lastpage{302}
\doi{10.1214/09-STS277REJ}
\referstodoi{10.1214/09-STS277}

\begin{document}
\begin{frontmatter}

\vspace*{6pt}

\title{Rejoinder: Likelihood Inference for Models with Unobservables Another View}
\runtitle{Rejoinder}

\begin{aug}
\author[a]{\fnms{Youngjo} \snm{Lee}\ead[label=e1]{youngjo@snu.ac.kr}\corref{}}
\and
\author[b]{\fnms{John A.} \snm{Nelder}\ead[label=e2]{j.nelder@imperial.ac.uk}}
\runauthor{Y. Lee and J. A. Nelder}

\affiliation{Seoul National University and Imperial College}

\address[a]{Youngjo Lee is a Professor, Department of Statistics,
Seoul National University, Seoul, Korea \printead{e1}.}
\address[b]{John A. Nelder is a Visiting Professor, Department of
Mathematics, Imperial College,
London, SW7 2AZ, UK \printead{e2}.}

\end{aug}



\end{frontmatter}

\section{Introduction}\label{sec1}

First we should like to thank the editor for allowing us to respond to
interesting discussions from the discussants, Molenberghs, Kenward and
Verbeke (MKV), Louis and Meng, for the effort they have put into their
replies, and for the many important points \ that they have raised.

We view statistics as comprising relationships between models and data,
where a statistical model is a formal mathematical formula which in some
sense represents the patterns in the data. It represents a tool underlying
the process of ``making sense of figures.'' There are two processes linking
models and data. The first, which we term the forward process, can be
written as
\[
\mbox{model}\longrightarrow \mbox{data}.
\]
This stands for, ``given a model, what would the data that it generates look
like?'' We call this process statistical modelling and it forms the basis of
probability theory. The second process, which we term the backward process,
can be written as
\[
\mbox{model}\longleftarrow \mbox{data}.
\]
This stands for, ``given data, and a (guessed) model what can we say about the
parameters in that\break model?'' We call this process statistical inference, and
it is displayed in Efron's (\citeyear{Efron98}) triangular diagram for 21st-century
statistical research, involving the three schools, Fisherian, Bayesian and
Frequentist. The process of inference involves two procedures, namely model
fitting and model checking. In the first we find values for the parameters
in the model that fit the data best, and in the second we use probability
theory to check whether the fit and, therefore, the assumed model is acceptable,
by looking at the distribution of a suitable badness-of-fit statistic. Model
checking could lead to a new model involving the two processes.

Among the discussants, MKV seem to suggest that data contain information
only about the parameters in the marginal likelihood, but not about the
unobservables (random effects) in the h-likelihood. Louis and Meng say that
extended likelihood such as h-likelihood does indeed carry information about
the unobservables, but that nevertheless the Bayesian approach is best
suited for such inferences. We hope to show how the ideas can be combined
together in the h-likelihood framework to give a new type of statistical
inference. We shall try to make clear the inferential status of our
framework.

The Bayesian framework is a well-defined mathematical structure about which
theorems can be proved. However, it requires a statement of subjective prior
belief about the unknown parameters which we are unable to provide. Of
course many attempts have been made to define ``objective'' priors, but we
believe them not to have been successful. As Barnard (\citeyear{Barnard95}) used to stress,
in scientific inference the aim is to look for objective conclusions that
scientists can agree upon. Senn (\citeyear{Senn08}) puts it more strongly when he writes,
``In fact the gloomy conclusion to which I am drawn on reading de Finetti
(\citeyear{deFinetti74}) is that ultimately the Bayesian theory is destructive of any form of
public statistics.'' An alternative description is that we are looking across
data sets for significant sameness, structures that remain unchanged when
external conditions vary, a view which has been strongly propagated by
Ehrenberg (\citeyear{Ehrenberg75}). Another problem of using priors on parameters is that
however many data are collected, no information is added regarding the
parameters in the prior. In contrast to the many possible priors in Bayesian
framework, in our system there is only one corresponding prior likelihood (Pawitan, \citeyear{Pawitan01})
for parameters, namely \mbox{$L(\theta )=1$}, and as data grow
information is accumulated on all parameters in the model. Model checking is
a vital part of inference and we regard accumulated information as necessary
for model checking to be effective. Science is an entirely open-ended
procedure, and there can be no possibility of assigning probabilities to all
the models we have not thought of yet.

Louis says that our paper is, ``more of an opinion piece than a scientific
comparison of approaches.'' In fact, we see no fault in presenting an opinion
piece; indeed, the title itself should prepare the reader for what follows.
We have indeed carried out extensive simulations with a wide variety of data
comparing our estimates with those of other methods, and so far h-likelihood
estimates have been often uniformly better in terms of mean-square error.
Some of these comparisons may be found in Lee, Nelder and Pawitan (\citeyear{LeeNelder06}). We
think our solution to mending the plug-in empirical Bayesian (Carlin and
Louis, \citeyear{Carlin00}) method is more straightforward than that of Louis which
requires yet another level of priors. We have used the Laplace approximation
when explicit expression for integration is not available. It seems to work
well in problems involving unobservables, except for a few extreme cases
where second-order Laplace approximations are required. We support the view
that, ``the art of applied mathematics is to know when to approximate.''

Louis seems to regard our paper as over-promotion of h-likelihood, but maybe
that is because we are highly enthusiastic about h-likelihood. In the paper
we do in fact make scientific comparisons as opposed to the typical Baysian,
who rarely makes comparisons with other approaches. As we shall discuss in
Section~\ref{sec4}, with a Bayesian model, the h-likelihood is equivalent to the Bayes
posterior so there is no disagreement in the case of a Bayesian model (see
Bj\o rnstad, \citeyear{Bjornstad96} for more discussion). We try to give a unified framework
for statistical methods developed by the three schools, but in our approach
we regard a fixed parameter as simply unknown so that it differs from
efforts of unification from the Bayesian side (Box, \citeyear{Box80}; Little, \citeyear{Little06}).

We shall now respond to the discussants by topic instead of responding to
individual contributions.

\section{Lp, Model Checking and Model Choice}\label{sec2}

The likelihood principle (Birnbaum, \citeyear{Birnbaum62}) provides a very good reason
for using likelihood for inferences, but it does not show how this should be
done. We have shown over the last 15 years how to do so. One drawback of LP
and likelihood methods in general is that they do not tell us what to do
when the model in use is not right. Suppose that we have a random-effect
model. If the model is right the standard sandwich standard error estimators
can be made from the marginal likelihood. However, we can have another
sandwich estimator from the h-likelihood which is useful when the
homogeneity assumption on the variance of random effects is violated (Lee, \citeyear{Lee02}).
H-likelihood also shows how to reduce the bias of the ML estimators
in frailty models with nonparametric baseline hazards (Ha, Noh and Lee, \citeyear{Ha09}).

MKV claim that inferences about unobservables cannot be made because they
are not identifiable. This is not so. Unobservables can change their status
to fixed unknowns once the sample has been observed, as we shall discuss in
Section \ref{sec6.2}. Thus in principle unobservables can be estimated and model
assumptions can be checked after the data have been collected. However,
unobservables occur in various forms, for example, as a random effect or as a
missing datum. One cannot extract any information about the latter from
observed data while this can be done about the former. Thus for the latter,
model checking based on observed data is not possible.

However, model checking is very important. We have shown that the
distributional assumptions on the random effect can be checked, and have
developed various model-checking procedures and criteria for model
selection. Suppose we have two different random-effect models,
\[
y_{i}=x_{i}\beta +W_{i}w+e_{i}\quad \mbox{and}\quad y_{i}=x_{i}\beta +U_{i}u+e_{i},
\]
which lead to the same marginal model. The two models could have different
numbers of random effects. For each model the assumptions about the random
effects can be checked by the model-checking procedures given in
Lee, Nelder and Pawitan (\citeyear{LeeNelderPawitan06}). If the assumed model is correct we can give estimates for
each random component. However, the individual random-effect estimators of
the two models cannot be matched. Nevertheless, Lee and Nelder (\citeyear{LeeNelder06}) show
that the two models, if equivalent, give the same inferences for equivalent
quantities, for example, that $W_{i}\hat{w}=U_{i}\hat{u},$ giving the same
predictions for the data, $\hat{y}_{i}(w)=x_{i}\hat{\beta}+W_{i}\hat{w}=%
\hat{y}_{i}(u)=x_{i}\hat{\beta}+U_{i}\hat{u}$. Now suppose that they lead to
the same marginal model for $y,$ but give different predictions for the
data, $\hat{y}_{i}(w)\neq \hat{y}_{i}(u).$ Then a model choice can be made
from the deviances
\[
D(w)=\sum \{y_{i}-\hat{y}_{i}(w)\}^{2}
\]
and
\[
D(u)=\sum \{y_{i}-%
\hat{y}_{i}(u)\}^{2},
\]
where $D(w)\neq D(u).$ These deviances are constructed from the conditional
likelihood $f_{\theta }(y|v)$ (see Lee and Nelder, \citeyear{Lee96} and Lee,
Nelder and Pawitan, \citeyear{LeeNelderPawitan06}, Chapter 6.5).
We, like Bayesians (Spigelhalter et al., \citeyear{Spiegelhalter02}) use the so-called deviance information criterion (DIC) based on its
degrees of freedom (see Ha, Lee and MacKenzie, \citeyear{Ha07} and Vaida and
Blanchard, \citeyear{Vaida05} for more discussion). MKV seem to regard two models as equivalent if
they lead to the same marginal model. How can the two models with different
predictions be equivalent? Consider a one-way random model (M1) leading to a
marginal multivariate model with composite symmetric covariance structure
(M2). They cannot be the same model because they give different predictors.
The former (M1) exploits the covariance structure to give a better
prediction. Regardless of how the data are generated, if the covariance
structure is composite a symmetric one-way random model gives a better
prediction than that based on the common mean of the marginal model (M2).
The random-effect model is in fact an advancement on the marginal model
because it shows how to predict. Various time series models and spatial
models have been proposed in this respect.

Suppose that the unobservables $v$ are missing data $y_{\mathrm{mis}}.$ Let $\hat{y}%
_{i}(R)$ and $\hat{y}_{i}(NR)$ be the predicted values of the missing data
under missing at random (MAR) and missing not at random (MNAR),
respectively. The deviances are then
\[
D(R)=\sum \{y_{i}-\hat{y}_{i}(R)\}^{2}
\]
and
\[
D(NR)=\sum \{y_{i}-%
\hat{y}_{i}(NR)\}^{2}.
\]
Now suppose that the two missing mechanisms, MAR and MNAR, give the same
predictions for the observed data while giving different predictions for
the missing data. Then we have
\[
D(R)=A+\sum \{y_{\mathrm{mis},i}-\hat{y}_{\mathrm{mis},i}(R)\}^{2}
\]
and
\[
D(NR)=A+\sum
\{y_{\mathrm{mis},i}-\hat{y}_{\mathrm{mis},i}(NR)\}^{2},
\]
where $A=\sum \{y_{\mathrm{obs},i}-\hat{y}_{\mathrm{obs},i}(R)\}^{2}=\sum \{y_{\mathrm{obs},i}-\break \hat{y}%
_{\mathrm{obs},i}(NR)\}^{2}$. In this case we cannot make a model choice based upon
the deviance because both $y_{\mathrm{mis},i}$ and their predictors are based upon
the model assumptions for the missing data. Even though\break $\hat{y}%
_{\mathrm{mis},i}(R)\neq \hat{y}_{\mathrm{mis},i}(NR)$ we cannot observe $y_{\mathrm{mis},i}$ to
evaluate them. In this case, given only the observed data, sensitivity
analysis can be used to show how inferences about $\hat{y}_{\mathrm{mis},i}(R)$ and $%
\hat{y}_{\mathrm{mis},i}(NR)$ vary as the model for the unobservables changes.
However, we may never be able to draw any conclusions from the analysis
because we do not have the data $y_{\mathrm{mis},i}$ to check our thinking. We are
also unable to say that we have checked for all the possible ways about
which our thoughts may be wrong.

We agree that care is necessary, as MKV say, but something can still be done
about inferences for random effects of the observed data.

\section{Nonparametric, Semi-parametric Models and GEE}\label{sec3}

It is not always easy to check all the assumptions of a given model. For
example, with binary data it is hard to check the distributional assumption
about the random effects. In semi-parametric frailty models with
nonparametric baseline hazards we can relax the specification of
probability models on certain parts of the model. However, this differs from
the lack of a probability basis, such as is shown in some GEEs. In a given
semi-/nonparametric model, there are many classes of submodels which belong
to the model. It is not correct that estimating equations such as the
quasi-likelihood estimating equations (father of GEE) satisfy only the first
two moment (or minimal) conditions; Lee and Nelder (\citeyear{Lee98}) showed that they
satisfy all the higher cumulant conditions of a GLM family if it exists for
the given mean and variance relationships. We have demonstrated that the
estimating equations implicitly impose assumptions about the higher
cumulants, so that a choice can be made depending upon the robustness of
model misspecifications. We are not\break against the use of GEE when it has a
proper model basis. However, its claimed advantage of less sensitivity to
model assumptions results from not comparing like with like.

In HGLMs, the distribution of random effects can be relaxed to give
nonparametric ML estimators\break (Laird, \citeyear{Laird78}). Parameter estimates from
binary\break
GLMMs can be sensitive to the distributional assumptions of the random
effects. A solution to this is to allow heavy-tailed distributions for the
random effects, to give a robust analysis (Noh, Pawitan and Lee, \citeyear{Noh05}). Thus
various parts of the model assumptions in HGLMs can be relaxed to produce
new nonparametric or semi-parametric models. Ha, Noh and Lee (\citeyear{Ha09})
show that in semi-parametric frailty models h-likelihood extends the
partial likelihood of Cox (\citeyear{Cox75}) to produce new efficient estimating
equations.

\section{APHLs versus Marginal Posteriors}\label{sec4}

Louis and Meng both say that our adjusted profile h-likelihood (APHL) in
Section 3.2 is a Laplace approximation to their marginal posterior. This is
not true because it can also eliminate fixed unknowns. Consider the
h-likelihood,
\begin{eqnarray*}
h&=&h(\theta ,v)=\log f_{\theta }(y|v)+\log f_{\theta }(v)=\log f_{\theta
}(y,v)\\
&=&\log f_{\theta }(y)+\log f_{\theta }(v|y).
\end{eqnarray*}
Bayesian models are composed of two objects, namely the data and
unobservables; $\theta $ is not fixed unknown, but unobservable. Thus their
model is
\[
B=h+\log \pi (\theta ),
\]
where $h=\log f(y|v,\theta )+\log f(v|\theta ).$ Thus the Bayes\-ian framework
eliminates $\theta $ by integration, even with the use of the improper prior
$\pi (\theta )=1.$

Suppose that all $\theta $ are indeed unobservables with a known
distribution $\pi (\theta )$. Then the extended LP says that $B$ carries all
the information about unobservables $(\theta ,v)$. In such a case the
Bayesian approach gives suitable statistical inferences, and the Bayesian
and h-likelihood inferences are equivalent.

But, suppose that $\theta $ represents fixed unknown parameters, rather than
unobservables. As Meng says there is no truly noninformative prior, at
least for continuous parameters. This means that for inferences about
parameters, use of the Fisherian likelihood would be suitable. To eliminate
nuisance parameters we could use profiling, conditioning or pivoting
methods, as developed by Fisherian and frequentist schools. Fisher (\citeyear{Fisher34})
shows that
\[
\log f_{\theta }(\hat{\theta}|A)\propto m(\theta )-m(\hat{\theta}),
\]
where $m(\theta )=\log f_{\theta }(y),$ $\hat{\theta}$\ is the ML estimator
and $A$ is an ancillary statistic. A wonderful generalization of Fisher's
work (Barndorff-Nielsen, \citeyear{BarndorffNielsen83}) gives the so-called magical formula,
\[
\log f_{\theta }(\hat{\theta}|A)\propto m(\theta )-m(\hat{\theta})+\tfrac{1}{2%
}D(m,\theta ),
\]
where $D(m,\theta )$ is defined in Section 3.2 of the main paper.

Let $\theta =(\xi ,\tau )$ with~$\xi $ being the nuisance parameters and~$%
\tau $ being the parameters of interest. Suppose that~$\xi $ and~$\tau $ are
orthogonal parameters. Because the ML estimator is asymptotically
sufficient, using the magical formula, we can eliminate the nuisance
parameter~$\xi $ from the marginal likelihood,
\[
\log f_{\theta }(y|\hat{\xi},A)=p_{\xi }(m;\tau ),
\]
where $p_{\xi }(m;\tau )$ is defined in Section 3.2 of the main paper,
giving the Cox-Reid (\citeyear{Cox87}) adjusted profile likelihood. This formula happens
to be the same as the Laplace approximation, integrating out $\xi .$ But it
is actually using the Fisherian method of conditioning out the fixed
parameters. This adjustment improves the profiling method (Lee, Nelder and Pawitan, \citeyear{LeeNelderPawitan06}).
We note that the elimination of parameters and unobservables can be
carried out in a uniform formula (APHL) which eliminates unobservables by
integration (as with the Bayesian approach) and parameters by conditioning
or (adjusted) profiling. Thus our APHL is quite different from the Bayesian
marginal posterior. We believe that the prior on fixed parameters is
informative if the APHL and marginal posteriors differ. In our framework we
use profiling, modified (or adjusted) profiling, or pivoting, as developed
by the likelihood school, to eliminate nuisance fixed parameters while we
use integrating-out techniques, as developed in the Bayesian school, to
eliminate unobservables. Our technique, among other things, extends REML
from normal models to the GLM class.

\section{How to Use the H-likelihood}\label{sec5}

Clearly Meng understands how to form the h-likelihood for inferences. Most
complaints are caused by misunderstandings about how to use the likelihood
for statistical inference.

\subsection{Neyman--Scott Problems and Nuisance Parameters}\label{sec5.1}

When the number of nuisance parameters increases with the sample size, the
ML estimators, jointly maximizing nuisance and parameters of interest
together, can give seriously biased estimates (Neyman and Scott, \citeyear{Neyman48}).
Similarly joint maximizations of the h-likelihood provide seriously biased
estimates, as Little and Rubin (\citeyear{Little83}) have shown. However, if we maximize an
appropriate APHL we can avoid such problems (Lee, Nelder and Pawitan, \citeyear{LeeNelderPawitan06}, Chapter 4).
When the number of nuisance parameters increases with the sample size,
the profile likelihood is often not satisfactory, and APHLs have been
developed for such cases. Yun, Lee and Kenward (\citeyear{Yun07}) showed that Little and
Rubin's (\citeyear{Little83}) complaints can be resolved by using appropriate APHLs.

Lee, Nelder and Pawitan (\citeyear{LeeNelder06}) highlight the main philosophical difference
between completed-data likelihood for the EM method and the h-likelihood. In
the former missing data are treated as unobserved data, while in the latter
they are nuisance parameters so that the technique developed for parameter
handling can be used for the efficient imputation of missing data (Kim, Lee and Oh, \citeyear{Kim06}). We have found that the h-likelihood may give very
good imputation even up to 95 percent missingness while the EM method
suffers from a slow convergence and distorted results with over 30 percent
missingness (Lee and Meng, \citeyear{LeeMeng05}). We believe that our methods will lead to
great improvements in the handling of missing data.

\subsection{Invariance for Parameter Estimation and Cauchy-Type
Distributions}\label{sec5.2}

MKV complain that the likelihood method gives point estimates with undefined
variance. Suppose that~$y$ follows an exponential distribution with the
log-likelihood $m=-\log \lambda -y/\lambda .$ The ML estimate for~$\lambda $
is $\hat{\lambda}=y.$ Here the observed Fisher information is $I(\hat{\lambda%
})=(-1/\hat{\lambda}^{2}+2y/\hat{\lambda}^{3})=1/\hat{\lambda}^{2}$ which
gives the correct variance estimator for $\hat{\lambda}.$ However, the ML
estimate of $1/\lambda $ is $1/y$ whose moment does not exist so that its
variance estimator from the likelihood theory could be seemingly
meaningless. The ML estimators are invariant with respect to any
transformation. When the sample sized is fixed, ML estimators may have scales
where their moments do not exist. Let $\theta =\delta /\varphi $ in
equation (20) of the MKV discussion. Then $y^{k}$ is the ML\ estimator for $%
\theta ^{k}$ with a finite variance, for example, when $0\leq k<1/2.$ The
concept of variance may not be a useful measure of uncertainty in the ML
estimator once the data are given. Whole likelihood curves or some feature
of it such as the curvature would be useful (Pawitan, \citeyear{Pawitan01}); see more
discussion below.

\subsection{Invariance for the Estimation of Unobservables and Exponential
Model}\label{sec5.3}

Meng points out that consistency theory may not be applicable to estimators
for unobservables. It has been one of the aims of the development of
h-likelihood procedure to overcome such difficulties. In our framework we
use the marginal ML estimators for fixed parameters whose invariance is
well established. To maintain such invariance for unobservables we fix the
scale of unobservables in defining the h-likelihood and use the mode for
inferences. The sample mean cannot maintain the invariance with respect to
transformation of unobservables while the mode does.\ Because we are using
the mode to derive point estimates for unobservables, its scale is important
in defining the h-likelihood if we are to have good inferential properties.
We appreciate that Meng gives a good theoretical contribution in deciding
that scale. Another advantage of the mode estimator over the sample mean is
that it also allows one to do model selection (Lee and Oh, \citeyear{LeeOh09}) instead of
model average. Recently, Ma and Jorgensen (\citeyear{Ma07}) have argued against the use
of mode estimates for random effects and proposed the use of the orthodox
best linear unbiased predictor (OBLUP) method. However, Lee and Ha (\citeyear{LeeHa09})
show that the h-likelihood mode estimation gives both statistically better
precision and maintains the stated level of coverage probability better than
the OBLUP method.

Consider the exponential model of Meng in Section~7 of his contribution for
predicting a future observation $u=y_{n+1}.$ In Section 7.2 he uses the
right h-likelihood $h(\lambda ,v;y)$ with $v=\log u,$ giving $\hat{\lambda}=%
\bar{y}_{n}.$ Note that
\[
\hat{u}(\lambda )=\mathrm{E}(u|y)=\lambda ,
\]
which is the so-called best predictor, giving $\hat{u}=\hat{u}(\hat{\lambda}%
)=\hat{\lambda}.$ \ Now compute the following information matrix as in (4.3):
\[
I(\hat{\lambda},\hat{u})=I(\lambda ,u)|_{\lambda =\hat{\lambda},u=\hat{u}%
}=\pmatrix{
\displaystyle\frac{n+1}{\hat{\lambda}^{2}} & -\displaystyle\frac{1}{\hat{\lambda}^{2}} \vspace*{2pt}\cr
-\displaystyle\frac{1}{\hat{\lambda}^{2}} & \displaystyle\frac{1}{\hat{\lambda}^{2}}}.
\]
Note that $\operatorname{var}_{\lambda }(u|y)=\operatorname{var}_{\lambda }(u)=$E($\hat{u}(\lambda
)-u)^{2}=\lambda ^{2}.$ Thus
\[
(-\partial ^{2}h/\partial u^{2}|_{\lambda =\hat{\lambda},u=\hat{u}})^{-1}=\biggl(%
\frac{1}{\hat{\lambda}^{2}}\biggr)^{-1}=\hat{\lambda}^{2}
\]
gives the correct estimate of $\operatorname{var}_{\lambda }(u|y).$  From
\[
I(\hat{\lambda},\hat{u})^{-1}=
\pmatrix{
\displaystyle\frac{\hat{\lambda}^{2}}{n} & \displaystyle\frac{\hat{\lambda}^{2}}{n} \vspace*{2pt}\cr
\displaystyle\frac{\hat{\lambda}^{2}}{n} & \displaystyle\frac{(1+n)\hat{\lambda}^{2}}{n}},
\]
we get the correct estimate of
\[
\operatorname{var}_{\lambda }(u-\hat{u})^{2}=\operatorname{var}_{\lambda }(y_{n+1}-\bar{y}%
_{n})^{2}=\frac{(1+n)\lambda ^{2}}{n}.
\]
This also gives the first-order approximation to
\begin{eqnarray*}
\operatorname{CMSE}(u) &=&\mathrm{E}\bigl\{\bigl(\hat{u}(\hat{\lambda})-u\bigr)\bigl(\hat{u}(\hat{\lambda}%
)-u\bigr)^{\prime }|y\bigr\} \\
&=&\operatorname{var}_{\lambda }(u-\hat{u}|y)^{2}=\lambda ^{2}+(\bar{y}_{n}-\lambda
)^{2}.
\end{eqnarray*}
The delta-method for $\hat{v}=\log \hat{u}$ gives what Meng has for the
variance of $\hat{v}$ but does not provide correct estimates for either $\operatorname{var}_{\lambda }(v-\hat{v})^{2}$
or $\operatorname{var}_{\lambda }(v-\hat{v}|y)^{2}.$

We have shown that there exists some analogy between inferences for the
fixed parameters and for unobservables. However, there also exist
differences between them. For a fixed constant, let say $\lambda =3,$ we
have $g(\lambda )=g(3)$ for any function $g(\cdot).$ Such an invariance is
meaningful for an estimator of an unknown constant and can be achievable by
consistency of ML estimators for fixed unknowns. Suppose that we are
estimating E$(u|y)$ of unobservable $u.$ Then, in general
\[
\mathrm{E}(g(u)|y)\neq g\{\mathrm{E}(u|y)\},
\]
with equality holding only if $g()$ is a linear function. Lee and Nelder
(\citeyear{LeeNelder05}) showed that maintaining invariance of inferences from the extended
likelihood for trivial re-expressions of the underlying unobservables leads
to the definition of the h-likelihood. Once the data are observed we treat
the unobservables such as random effects as fixed unknowns as discussed in
the next Section so that we maintain $g(\hat{u})=\widehat{g(u)}$ for any
function $g(\cdot).$ However, the interpretation of the maximum h-likelihood
estimator as $\widehat{\mathrm{E}(u|y)}$ holds only on a particular scale $u$.

\subsection{Invariance for Parameterization and ML Estimation}\label{sec5.4}

In Bayarri's example, $y^{k}$ is the ML estimator of~$\tau $ for $\tau
=\theta ^{k}.$ Here the consistency of $\hat{\theta}^{k}$ fails, so that the
unbiasedness of $\hat{\theta}^{k}$ and exact variance estimator of $\hat{%
\theta}^{k}$\ would be meaningful properties to achieve. Note that E($\hat{%
\theta}^{k})$ becomes $\theta ^{k}$, and $\widehat{\operatorname{var}(\hat{\theta}%
^{k})}=I(\hat{\theta}^{k})^{-1}=(-\partial ^{2}m/\partial \tau ^{2}|_{\tau =%
\hat{\tau}})^{-1}=2k^{2}\hat{\theta}^{2k}$ becomes $\operatorname{var}(\hat{\theta}^{k})$
as $k$ approaches zero. Thus these desirable properties can be achievable
on a particular scale of $\theta $. The existence of such a scale is
important for inferences about fixed unknowns. For example, if there exists
an exact confidence interval for a particular scale of a fixed parameter
[let's say ($\hat{\tau}-L,\hat{\tau}+U)]$, then it allows exact intervals for
all parameterizations of $\tau $\ (($g(\hat{\tau}-L),g(\hat{\tau}+U))$ for
any function $g(\cdot)$. A plot of the whole likelihood curve is useful to find a
proper scale of intervals for unobservables too (Lee and Ha, \citeyear{LeeHa09}).

\section{Is the H-likelihood Interval for Unobservables a Credible or
Fiducial or Confidence Interval?}\label{sec6}

Suppose that we have a model for the three objects ($y,\theta ,v)$ where $y$
and $v$ are random variables (RVs) and~$\theta $ is a fixed unknown
parameter. The statistical model $f_{\theta }(v)f_{\theta }(y|v)$ describes
how the RVs $y$ and $v$ are generated.

Once the data are observed as $y_{o,}$ where the subscript $o$ stands for
``observed,'' $y_{o}$ are fixed knowns, no longer RVs. In the Bayesian
framework all unknowns are considered as random without allowing fixed
unknowns. Bayesians assume a prior for $\theta ,$ making them random
variables (RVs) so that the marginal posteriors $\pi (\theta _{1}|y_{o}),$ $%
\pi (v_{1}|y_{o}), \ldots$ etc. can be obtained by integrating out the rest of
unknowns, regardless of whether these are fixed unknowns or unobservables.
We find it mysterious how the fixed unknown parameters $\theta $ can change
their status to RVs, leading to a prior probability $\pi (\theta )$ that
generates $\theta $.

\subsection{Intervals for Parameters}\label{sec6.1}

Based on the likelihood $L_{y_{o}}(\theta )=f_{\theta }(y_{o}),$
frequentists can derive confidence intervals for $\theta $ (random interval
for fixed unknown), and this has become a standard procedure. However, their
argument is based on repeated sampling from the same population (RSSP) to
which Fisher objected strongly. According to Fisher, when a scientist
seriously ``repeats'' an experiment he always has in mind at least the
possibility that the population of the previous experiment may turn out not
to be the same population from which he is currently sampling. If he really
knew the population of the new experiment was exactly the same as before he
would think of himself either as enlarging his original experiment or
wasting his time. So Fisher believed that the repeated sampling from
different populations (RSDP) was crucial in making intervals for unknowns.
The Bayesian credible interval, based upon $\pi (v_{1}|y_{o}),$ meets this
goal because inferences are confined to the experiment of the observed data $%
y_{o}$. Fisher tried to make an interval for $\theta $, the so-called
fiducial interval, under RSDP. But this RSDP assumption is too strong a
requirement for his fiducial interval to be applicable in general
(Barnard, \citeyear{Barnard95}) while the confidence interval can be made in various contexts.

\subsection{Intervals for Unobservables}\label{sec6.2}

The EB interval based on $f(v_{1}|y_{o},\hat{\theta})$ has different
treatments for fixed unknowns $\theta $ and unobservables $v.$ However, it
cannot account for the information loss caused by estimating $\theta ,$
leading to very liberal intervals. The Bayesian credible interval based on $%
\pi (v_{1}|y_{o}),$ while improving the EB interval a lot, can still exhibit
strange behavior as shown in Figure 4 of the main paper. Louis has suggested
some other priors to try. However, we object to putting priors for fixed
unknowns and recommend using likelihood methods for inferences about them.

The observed data can be obtained as follows: From the statistical model $%
f_{\theta }(v)$ the unobservables are realized (generated) as $v_{r}$ where
the subscript $r$ stands for ``realized.'' Note the use of the term
``realized'' instead of ``observed'' to emphasize that they are fixed but still
unknown. Then, the observed data $y_{o}$ are obtained from the model $%
f_{\theta }(y|v_{r}).$ Now suppose that we want to make an interval for the
fixed unknowns $v_{r}$ given $y_{o}.$

For linear mixed models, Henderson (\citeyear{Henderson75}) shows that the standard error
estimate from the Hessian matrix $I(\beta ,v)$ in (4.3) of the main paper
gives an estimate of the unconditional MSE
\[
\mathrm{E}\bigl\{\bigl(\hat{v}(\hat{\theta})-v\bigr)\bigl(\hat{v}(\hat{\theta})-v\bigr)^{\prime }\bigr\}.
\]
In 1996, we showed that this holds more generally in HGLMs. This means that
our proposed interval can be viewed as a confidence interval (random
interval for fixed unknown) for RSSP whose probability statement is for
unobserved future data. Simulation results in Section 4.3 are from RSSP. The
proposed 95 percent interval may not always cover $v_{r}$, but among 100
intervals, 95 of them are expected to cover the realized value.

It was Booth and Hobert (\citeyear{Booth98}) who showed that $I(\beta ,v)$ can also give
an estimate of CMSE$(v)=\break\mathrm{E}\{(\hat{v}(\hat{\theta})-v)(\hat{v}(\hat{\theta}%
)-v)^{\prime }|y_{0}\}$ for GLMMs with independent random effects. This
result can be extended to nonnormal random-effect models (Lee and Ha, \citeyear{LeeHa09})
and the correlated random-effect models in Section~4.3.1
(Lee, Jang and Lee, \citeyear{LeeJang09}). Given the observed data, we can therefore make an interval (fixed
interval for fixed unknown) whose probability statement is for all possible
future realizations of $v$. Fisher's aim of making intervals for RSDP may be
generally achievable for realized values of unobservables.

Louis says that we cannot account for the information loss caused by
estimating the dispersion parameters while Bayesian marginalization can do
so. In HGLMs dispersion parameters are orthogonal to the rest of parameters
so that we do not actually need to account for the information loss caused
by estimating them. In general, exactly the same method, that is, $%
I(\theta ,v),$ is used to account for estimating all the parameters. The
resulting interval is also identical to Kass and Steffey's (\citeyear{Kass89})
approximate Bayesian credible interval (fixed interval for random unknown)
with $\pi (\theta )=1$ (Lee, Jang and Lee, \citeyear{LeeJang09}).

The probability statement of Bayesian interval\break based on $\pi (v_{1}|y_{o})$
is different from the previous two intervals for fixed unknown. Probability
statement of Bayesian credible interval is for RV, that is, the 95 percent
Bayesian interval contains the unobservable with 95 percent probability
given the data. Such a probability statement may not be relevant to the
realized values of unobservables, but it is meaningful for inferences about
unrealized unobservables, for example, inferences about future unobserved
observations. Our proposed interval also allows such a statement for future
observations without requiring priors on $\theta $.

Our interval for unobservables could be a fiducial (fixed interval for fixed
unknown), frequentist (random interval for fixed unknown) or Bayesian
interval (fixed interval for random unknown), so allowing three different
interpretations. Similarly, the APHL can be interpreted either as an
approximate conditional likelihood eliminating nuisance fixed parameters by
the magical formula or as an approximate marginal posterior using the
Laplace approximation with $\pi (\theta )=1$.

Louis says for nonstandard problems where the purpose of analysis is to
find the shape of the distribution for the unobservables ($\pi (v_{1}|y_{o})$%
), Bayesians can offer a better algorithm using the MCMC method. For such
problems we agree that MCMC-type methods are useful, but the question is
whether we can do this without a subjective prior. Can we find a solution on
which all three schools can agree? Why not consider MCMC applied to
h-likelihood?

\subsection{Asymptotic versus Finite Sample Properties of ML
Estimators}\label{sec6.3}

Justification of ML inferences has relied heavily on asymptotic theory. On
the other hand, Bayesian inferences are exact in finite samples, but require
the B-club fee, priors for fixed unknowns. However, agreed or agreeable
priors may hardly exist. We have illustrated that an exact finite-sample
solution for the ML method is possible by finding a proper scale, but that
search for an exact scale cannot be an easy task. However, a practically
satisfactory scale can be often found without much difficulty and simulation
results in Section 4.3 of the main paper show that likelihood inference
using such an approximate scale can give a better finite-sampling property
than putting unjustifiable priors for fixed unknowns.\break Therefore, there is a
way of deriving finite-sample likelihood inferences without paying the
B-club fee.

\section{Three in One}\label{sec7}

Likelihood inferences are for models with two objects, namely, the data and
fixed unknowns (parameters), while the probability-based inferences of
Bayesians are for models with data and random unknowns (unobservables).
Statistical models of recent interest often have both parameters and
unobservables. So it seems beneficial to combine the inference methods
developed by the three schools. We are glad that Meng found pivoting to be
easy and useful for eliminating $\theta $ for his Bayesian inferences.
Frequentists use probability statements to evaluate the performance of
statistical methods. They use unobservables (unobserved future data obtained
by RSSP) to invoke probability statements. By allowing all three objects in
statistical models and inferences, we hope to accommodate the advantages of
all three schools in a unified framework.

In discussing Lee and Nelder (\citeyear{Lee96}), Smith wondered whether it was time to
bring the ``two cultures'' together. In Bayes's original paper (Bayes, \citeyear{Bayes63})
an example is given of balls rolled on a table; this seems to us to be
naturally expressible as a two-stage likelihood problem, leading to the
question, ``Was Bayes a Bayesian in the modern sense?''  We are very aware that
the model class we consider, though having a wide scope, is as yet
incomplete, and also that there are aspects of the theory, in particular on
the choice of scale for unobservables in the definition of the h-likelihood
and on the choice of scale for exact finite-sample likelihood inferences,
which are as yet incomplete. We are, however, particularly excited by the
contribution from Meng which seeks to connect our procedures to other
generally accepted statistical ideas. We hope that the time has indeed come
to bring the three cultures together!

\end{document}